\documentclass[aps,pre,superscriptaddress,twocolumn,showpacs,floatfix]{revtex4-1}
\usepackage{amsmath,amssymb,latexsym}

\usepackage{graphicx}

\begin{document}

\title
{Theory of complex fluids in the warm-dense-matter regime, and 
application to an unusual phase-transition in liquid carbon. 
}

\author
{
 M.W.C. Dharma-wardana}
\affiliation{
National Research Council of Canada, Ottawa, Canada, K1A 0R6
}
\email[Email address:\ ]{chandre.dharma-wardana@nrc-cnrc.gc.ca}

%
\date{\today}

\begin{abstract}
Data from recent laser-shock experiments,   density-functional theory (DFT)
with molecular-dynamics (MD), and  path-integral Monte Carlo (PIMC)
simulations on carbon are compared with predictions from the
neutral-pseudo-atom (NPA)+ hyper-netted-chain (HNC)  approach for carbon, a
complex liquid in the warm-dense matter regime. The NPA results are in good
agreement, not only with high-density regimes that have been studies via
PIMC, but even at low densities and low temperatures where transient
covalent bonding dominates ionic correlations. Thus the `pre-peak' due to
the C-C bond at $\sim$1.4-1.6 \AA$\,$ and other features found in the
pair-distribution function from DFT+MD simulations   at 0.86 eV and 3.7
g/cm$^3$ etc., are recovered accurately in the NPA+HNC calculations. Such
C-C bonding peaks have not been captured via average-atom ion-sphere (IS) models.
Evidence for an unusual liquid $\to$ vapor and metal$\to$ semi-metal
transition  occurring simultaneously is presented. Here  a
strongly correlated metallic-liquid with transient C-C bonds, i.e., carbon
at density $\sim$ 1.0 g/cm$^3$ and mean ionization $Z=4$ transits abruptly
to a disordered mono-atomic vapour  at 7 eV, with $Z\simeq$ 3. Other
cases where  $Z$ drops abruptly are also noted. The nature of $Z$, its
discontinuities, and the role of exchange-correlation, are
reviewed.  The limitations of IS models in capturing the physics
of transient covalent bonding in warm dense matter are discussed. 
\end{abstract}

%
\maketitle
\section{Introduction.}
\label{intro} 
A number of light elements like hydrogen, boron, carbon, nitrogen,
 silicon, phosphorous, etc., and their mixtures form strong
 covalent bonds in the solid and retain much of
this bonding, even when molten~\cite{galli89,glosli99,
ghiring05} and in warm-dense-matter (WDM) regimes~\cite{whitley15, kraus13,
hammelCH12, DWP-Carb90}.  A theory for predicting the properties of these
elements and their mixtures reliably and rapidly is needed  in many
technological applications. Their properties under higher compressions and
temperatures are  also of great interest in
astrophysics~\cite{hubNellis91,sherman12,driver12}.
Carbon is an important component of  inter-stellar matter,
white dwarfs, solar and extra-solar planets.
WDM carbon is basic to many technologies and 
 inertial-confinement fusion (ICF) ablators~\cite{lindl04,benedict2014}. 

Many  applications require the equation of state (EOS) and transport
properties  in experimentally inaccessible regimes. The experiments
when feasible are quite demanding ~\cite{kraus13,savvati08}.  The
theoretical prediction requires the electronic and ionic structure factors,
and their interaction potentials. At low temperatures (compared to the Fermi
energy $E_F$), numerically expensive density-functional theory (DFT)
calculations  coupled to molecular-dynamics (MD)  provide a reliable method
if sufficiently large calculations can be made using suitable
exchange-correlation (XC) functionals.  In DFT the WDM is modeled by a
sequence of quenched $N$-atom periodic crystals thermally evolved using MD.
The spurious electronic  band structure of each crystal is averaged over the
many ionic configurations obtained from MD. Thus DFT+MD  is  impractical at
higher-temperatures ($T$)  due to the many electronic states needed. At high
$T$ and high density ($\bar{\rho}$) path-integral Monte Carlo (PIMC) methods
are available for small-$N$ (e.g, 24 atoms) simulations~\cite{driver12}.

 Since the number of ions $N$ in DFT simulations is limited, other 
methods are used, especially for ambient-temperature applications.
Semi-empirical methods  developed from the ``embedded-atom'' approach have led
to ``reactive-potentials''  which include bonding, dependence on coordination, torsion
and bond angle.  Carbon is a complex liquid where conjugated bonds
($sp3, sp2, sp$) and weak graphite-like partial conjugation occur
deploying four valance electrons, i.e. with mean ionization $Z$=4. The Brenner
potentials incorporate conjugation and were used by Glosli {\it et
al.,}~\cite{glosli99} in their study of the phase transition in liquid carbon.
The modern bond-order potentials (LRBOP)  by  Los {\it et
al.}~\cite{losFaso03,losGhir05}  include
long-range effects and are fitted to a very wide data base.
 However, after applying LRBOP to calculate $S(k)$ of liquid C at
0.52 eV, Kraus {\it et al.}~\cite{kraus13} stated that ``this potential appears
to be too stiff. Compared to DFT+MD, we find a higher pressure and a more
pronounced structure which is not found in the experiment''.  Thus LRBOP is
hardly promising for  $T\sim E_F$ typical of WDM since $T$=0.52 eV,
$T/E_F=0.017$, defines a very low-$T$ WDM where it is already inadequate.

Here we use a first-principles  DFT approach treating electrons
with a
one-body density $n(r)$, and ions  with a one-body density $\rho(r)$ 
to present  results using the neutral-pseudo-atom  (NPA) model.
 This couples
one-body Kohn-Sham calculations and its electron-ion
 potentials to  pair-distribution functions (PDFs) via integral equations
like the hyper-netted-chain (HNC) equation, or MD simulations.
NPA+HNC closely recovers PIMC results, and DFT+MD results including C-C
covalent bonding signatures in the ion-ion PDFs at low -$T$, but extends seamlessly
 to high $T$ regimes inaccessible by DFT+MD. In Sec.~\ref{monomer-dimer.sec}
we present evidence for  a phase transition of highly-correlated
liquid carbon containing transient C-C bonds into a weakly correlated
mono-atomic carbon gas. The  pressure,  compressibility and conductivity in the
neighborhood of the transition are presented. The  nature of the mean ionization
$Z$, its behaviour near the transition, and comparisons of the NPA+HNC method
with PIMC and DFT+MD are given. An appendix discusses details of the electronic
structure near the phase transition, and the nature and constraints on the mean
ionization $Z$.
\section{The neutral-pseudo-atom model in the context of complex fluids.}
\label{NPA.sec}
Here we demonstrate that the NPA approach~\cite{cdw-cpp15,Pe-Be,PDWBenage02,eos95}
 works sufficiently well for complex liquids in their 
{\it metallic phase}, even near the melting point, as  already claimed by
the present author and  Perrot in 1990~\cite{DWP-Carb90} by comparing
NPA-type calculations with DFT+MD calculations of Galli {\it et
al.}~\cite{galli89} for carbon. There are two types of atom-in-plasma models, viz.,
Ion-Sphere (IS) models, e.g,~\cite{Murillo13,Muze05,Purgatorio}, and the
Correlation-Sphere (CS) model used by Perrot and Dharma-wardana, as well
 as some possible
intermediate models.  We show that the NPA model used here captures the C-C
bonding effects  not  captured so far in IS models of various
sorts, as  explicitly stated by, e.g., Starrett
 {\it et al.}~\cite{StaSauDalHam14}. Furthermore, we discuss the 
evaluation of the mean electron density per
ion, viz., $Z$ since Blenski {\it et
al.}~\cite{Blenski2013, PironBlenski11} and Stern {\it et al.}~\cite{SternZbar07}  have pointed
out difficulties in their models in defining $Z$.

In NPA a nucleus of charge $Z_n$ is immersed in its electron
distribution $n(r)$ and the corresponding ion distribution $\rho(r)$, where
each `field'  ion carries its shell of bound electrons compactly contained
within their  Wigner-Seitz (WS) spheres. The mean electron- and ion- 
densities become   $\bar{n}$ and $\bar{\rho}$ far away from the nucleus, say
at $r\ge R_c$. This is taken as the radius of the `correlation sphere' (CS) which
is large enough to be an  effective `infinite volume'. Defining the
Wigner-Seitz radius  $r_{ws}=(3/4\pi\bar{\rho})^{1/3}$ of the ions, we use $R_c$
$\sim 10r_{ws}$ for strongly coupled systems, while  $R_c \sim 5
r_{ws}$ may be enough  for weakly coupled systems, e.g., at high $T$. The IS
model is obtained if $R_c$ is curtailed to $r_{ws}$. The Kohn-Sham and HNC
equations are solved within this large CS. The $R_c$ plays the
role of the linear dimension of the DFT+MD simulation box.
However, in DFT+MD, if $N\sim 100$, the  equivalent linear dimension is
$\simeq 2.3 r_{ws}$, while the $R_c\simeq 10r_{ws}$ in the NPA+HNC mimics a
1000-atom volume.

DFT  asserts that the free energy $F([n,\rho])$ is a functional of the {\it
one-body} electron density $n(r)$, and the {\it one-body} ion density
$\rho(r)$, irrespective of complex interactions (e.g., superconductive
associations for electrons), and complex-bonding among ions.
DFT constructs  non-interacting ``Kohn-Sham
electrons'' that move  in the one-body Kohn-Sham potential, and similarly
there are non-interacting ``Kohn-Sham'' ions moving in the ``potential of
mean-force'' $V_m(r)$. The latter is the classical  DFT potential acting
on an ion, well known in the theory of classical fluids. The
variational problem  invokes a {\it coupled pair} of DFT 
equations~\cite{DWP1}.
\begin{eqnarray}
\label{F-elecden.eqn}
\delta \Omega([n],[\rho])/\delta n &=& 0 \\
\label{F-ionden.eqn}
\delta \Omega([n],[\rho])/\delta \rho &=& 0
\end{eqnarray}
The functional derivatives of the grand potential $\Omega$ are implied. The
first equation leads to a Kohn-Sham equation which invokes  an
exchange-correlation functional $F^{ee}_{xc}(n)$ for the electrons in the
potential created by the ions, while the second variationally
determines the DFT-ion distribution $\rho(\vec{r})$  and invokes an ion-ion 
correlation functional  $F^{ii}_{c}(\rho)$. The ions,  screened by the
electron distribution  are classical particles without exchange  (see
Eq.~1.13, p 628 Ref.~\onlinecite{ilciacco93}). The equations
~(\ref{F-elecden.eqn}), (\ref{F-ionden.eqn})  are connected by a Lagrange
multiplier $Z$  enforcing the charge neutrality of the system,
$\bar{n}=Z\bar{\rho}$, as in Ref.~\cite{DWP1}. Note that both
$F^{ee}_{xc}(n,\rho)$ and $F^{ii}_{c}(n,\rho)$ contain cross terms of the
form $F^{ei}_{xc}(n,\rho)$ arising from the electron-ion contribution to the
grand potential, viz.,  $\Omega_{ei}(n,\rho)$. These are discussed by
Dharma-wardana in Ref.~\cite{ilciacco93}, by
Chihara~\cite{Chihara87}, and  by Furutani {\it et al.}~\cite{Furutani90}.  

The usual DFT+MD codes do not use Eq.~\ref{F-ionden.eqn}, or an
$F^{ii}_c([\rho])$ since the positions of the $N$ ions are explicitly in the
simulation and held fixed by invoking the Born-Oppenheimer approximation
(BOA). The BOA is not needed in the NPA. The use of $\rho(r)$ leads to major
  simplifications since, instead
of having an $N$-center problem, we have a one-body problem which can be
reduced to a `single-ion problem'. But the multi-ion correlations captured
by the $N$-ion problem have to be picked up via the $F^{ii}_c[\rho]$.
Given the $F^{ii}_c$, this finite-$T$ `average-atom' defined by
Eqs.~(\ref{F-elecden.eqn}), (\ref{F-ionden.eqn}) is a formally exact DFT
concept  extending the usual Kohn-Sham formalism.
Approximations arise in formulating $F^{ee}_{xc}$ for
electrons, and in constructing $F^{ii}_c$ for ions.  In
Refs.~\cite{DWP1,ilciacco93} $F^{ii}_c$ was shown to be a sum of
hyper-netted-chain (HNC) + bridge diagrams and this is known as the MHNC
model used here when appropriate. Given  $m$ species of ions,
e.g., several ionization states, or   molecule formation, then a matrix
of such correlation functionals $F^{ij}_c(\rho_1,\cdots,\rho_m)$ are needed
 as in multi-component HNC theory~\cite{eos95}.

We use a one-component treatment here since the complex fluids
studied are in a regime where {\it no stable} molecules are formed,
while transient molecules may formed. Spherical symmetry around any nucleus
is assumed, as appropriate for a fluid  WDM, while the method can  also
be used for  solids~\cite{Dagens2,HarbourCCP15}. The optional  assumption of
radial symmetry gains  simplicity while sacrificing lattice-like
information important only at very low-$T$. Another useful
step is to decouple Eq.~\ref{F-elecden.eqn} and Eq.~\ref{F-ionden.eqn}
instead of the concurrent solution used in Ref.~\onlinecite{DWP1}.  A
simple approximation to the ion distribution $\rho(r)$ is taken to be a
``spherical cavity'' of radius $r=r_{ws}$ with $\rho=\bar{\rho}$ for
$r>r_{ws}$. Here $r_{ws}=(3/4\pi\bar{\rho})^{1/3}$ is the Wigner-Seitz (WS)
radius of the ions.  Unlike in most other AA models, this cavity-like ion
distribution is a calculational device  whose effect will be  subtracted
out. The cavity with the nucleus at its center creates an object
with net zero charge i.e., a weak scatterer. The effect of the WS-cavity in
$\rho(r)$ on the electron subsystem of density $\bar{n}$ is assumed to be
given  by finite-$T$ linear response theory.

The above  approximations are justified for the low $Z_n$ elements where
the bound-electron core of radius $r_c$  is compactly contained in the
WS-sphere with $r_c<r_{ws}$, unambiguously defining the number of bound
electrons $n_b$ and an effective ionic charge $Z=Z_n-n_b$. 
 The number of bound electrons is
calculated from the Fermi occupation factors $f_{nl}(\epsilon_{nl},T)$ of
the Kohn-Sham bound states (quantum numbers $n,l$), with eigenvalues
$\epsilon_{nl}<0$. Kohn-Sham theory uses a {\it homogeneous}
non-interacting distribution of electrons at the interacting density
$\bar{n}$ as the reference state, and the chemical potential is the
non-interacting value $\mu^0$, while
$f_{nl}=1/\left[1+\exp\{(\epsilon_{nl}-\mu^0)/T\}\right]$.  In the NPA, the
plasma outside the correlation sphere ($r>R_c$) is bulk-like. Unlike in
ion-sphere models, the
Kohn-Sham potential for $r>R_c$ is the NPA reference zero potential
(see Eq. 2.11 of Ref.~\cite{DWP1}); hence
 the effective $\mu$ is simply $\mu^0$. The $Z$ defined via the bound states
is
\begin{equation}
\label{nb-Z.eq}
Z=Z_b=Z-n_b;\, n_b=\sum_{nl}2(2l+1)\int d\vec{r} f_{nl}|\phi_{nl}(r)|^2
\end{equation}
Here $\phi_{nl}(r)$ is a bound-state (i.e, $\epsilon_{nl}<0$) solution of the
Kohn-Sham equation.  

 The NPA
model satisfies the finite-$T$  Friedel sum rule which applies to the
plane-wave like continuum states
$\psi_{kl}(r)$, $\epsilon_k=k^2/2$ and having phase shifts $\delta_l(k)$. 
\begin{equation}
\label{Friedel.eq}
Z=Z_F=\frac{2}{\pi T}\int_0^\infty kf(\epsilon_k)\{1-f(\epsilon_k)\}
 \sum_l(2l+1)\delta_l(k)dk
\end{equation}
Thus we see that the Kohn-Sham solution for the
bound and continuous spectra, and $Z$ are
tightly controlled by the Friedel sum rule,  the $f$-sum rule, and charge
neutrality. Charge neutrality implies that
\begin{equation}
\label{cn.eq} 
Z=Z_{cn}=\bar{n}/\bar{\rho}=(4\pi/3)\bar{n}r_{ws}^3
\end{equation}
We need $Z=Z_b=Z_F=Z_{cn}$ at self-consistency.  The
NPA consists of the nucleus surrounded by its 
electron distribution $n(r)$ and the input ion distribution  $\rho(r)$
extending up to $r=R_c$ forming a neutral object. The presence of
the cavity (or $\rho(r)$) makes the net NPA charge to be zero, and the
 Friedel sum is zero precisely for that WS-cavity whose charge
  $Z_{cn} =Z$. Hence
a trial electron density $\bar{n}$ is input and  the ion
density $\bar{\rho}$ (i.e., cavity radius $r_{ws}$) which self-consistently
 satisfies the neutrality and sum rules etc., is calculated. That $\bar{n}$
 which yields the required physical $\rho$ is determined iteratively.

Note that $n(r)$
extends over the  whole correlation sphere $r=R_c$, and not limited to
the WS-sphere, $r=r_{ws}$ as used in many `atom-in-plasma' (AIP) models.
Furthermore, such AIP models where the ``pseudo-ion'' is restricted to an
ion sphere do not have a Friedel sum rule, and cannot use the
non-interacting $\mu^0$ as required by DFT. This is discussed more
fully in Sec~\ref{DisconZ}.

The so-called
`Chihara decomposition' separating bound and free spectra of an ion,
invoked in expressing one-atom properties,  e.g., in the
`ion-feature' needed for X-ray Thomson scattering (XRTS) theory~\cite{GlenzerRedmer09},
occurs naturally in the NPA . However, the  Kohn-Sham states do not
correspond to the physical one-particle states of the electrons. They are
given by the corresponding Dyson  equation~\cite{PDW-levelWidth}. Although
the NPA is not a physical object, it is a rigorous DFT construct. The NPA
becomes thermodynamically variational when it is coupled with
Eq.~(\ref{F-ionden.eqn}). 
\subsection{Calculation of pair-potentials and 
the ion distribution $\rho(r)$.} 
In most cases the above simplifications leading to the NPA hold, and
the electron distribution $n(r)$ around a nucleus in the simplified ion (i.e., a cavity)
distribution is easily calculated from the Kohn-Sham equation.  The next
step is to solve  Eq.~(\ref{F-ionden.eqn}) for the accurate ion
distribution $\rho(r)=\bar{\rho}g_{ii}(r)$ where $g_{ii}(r)$ is the
ion-ion PDF. In Ref.~\cite{DWP1} we show that solving Eq.~(\ref{F-ionden.eqn})
is equivalent to solving an HNC-type equation. Thus $\rho(r)$ is
 evaluated from the ion-ion pair-potential determined
entirely by the electron charge density $n(r)$, the input nuclear charge
$Z_n$ and temperature $T$. The  electron density $n(r)$ of the NPA  is
corrected for the presence of the spherical cavity using perturbation
theory. Then the resulting $n(r)$ is written in the form: 
\begin{equation}
\label{n-division.eq}
n(r)=n_b(r)+\Delta n_f(r)+\bar{n}
\end{equation}
This division is unambiguous  since the bound density $n_b(r)$ is
compactly inside the WS sphere.   Since $\bar{n}=Z\bar{\rho}$, and since
$\bar{\rho}$ is the given matter density, $Z$ is simply the free-electron
density per ion in the plasma. However, some authors have claimed that $Z$
``..does not correspond to any well-defined observable in the sense of
quantum  mechanics'', i.e, that there is no quantum operator corresponding
to $Z$~\cite{PironBlenski11}. Within that view, there is no quantum
 operator for the
temperature or the chemical potential either.  We  review AIP
models of $Z$ and its physical nature in more detail in sec.~\ref{DisconZ}.
We subscribe to the standard view that $Z$ is a measurable physical quantity.

The Fourier transform of the cavity-corrected  free-electron pileup, viz., 
$\Delta n_f(k)$ is used to construct an electron-ion pseudopotential
$U_{ei}(k)$. We omit various technical details of the NPA model as they have
been amply discussed in many publications since the 1970s in applications
 to $T=0$ metals, finite-$T$ plasmas,
~\cite{Dagens2,DWP1,Pe-Be,eos95,PDWBenage02} and more
 recently~\cite{cdw-Utah12,Murillo13,xrt-LH16}
for applications to ultra-fast two-temperature
matter. The cavity corrected $\Delta n_f(k)$ is the  density modification
occurring in a {\it uniform} electron gas with a static neutralizing
background, arising from an ion of charge $Z$ placed in it.  It  determines 
$U_{ei}(k)$ and the ion-ion pair potentials $V_{ii}(k)$, via
the fully-interacting static-electron response function $\chi(k,T_e)$. In
Hartree atomic units,
\begin{eqnarray}
\label{pseudo.eq}
U_{ei}(k) &=& \Delta n_f(k)/\chi(k,T_e),\\
\label{resp.eq}
\chi(k,T_e)&=&\frac{\chi_0(k,T_e)}{1-V_k(1-G_k)\chi_0(k,T_e)},\\
\label{lfc.eq}
G_k &=& (1-\kappa_0/\kappa)(k/k_\text{TF});\quad V_k =4\pi/k^2,\\
\label{ktf.eq}
k_{\text{TF}}&=&\{4/(\pi \alpha r_s)\}^{1/2};\quad \alpha=(4/9\pi)^{1/3},\\
\label{vii.eq}
 V_{ii}(k) &=& Z^2V_k + |U_{ei}(k)|^2\chi_{ee}(k,T_e).
\end{eqnarray}
Here $\chi_0$ is the finite-$T$ Lindhard function, $V_k$ is the bare Coulomb
potential and $G_k$ is a local-field correction (LFC). The finite-$T$
compressibility sum rule for electrons is satisfied since $\kappa_0$ and
$\kappa$ are the non-interacting and interacting electron compressibilities
respectively, with  $\kappa$ matched to the $F_{xc}(T)$ used in the Kohn-Sham
calculation. In Eq.~\ref{ktf.eq}, $k_\text{TF}$ appearing in the LFC is the
Thomas-Fermi wavevector. We use a $G_k$ evaluated at $k\to 0$ for all $k$
instead of the more general $k$-dependent form (e.g., Eq.~50  in
Ref.~\cite{PDWXC}) since the $k$-dispersion in $G_k$ has negligible effect
for the WDMs of this study.

Eq.~\ref{vii.eq} gives the pair-potential to second order in the
pseudopotential. Higher order terms can become important, as
displayed in Fig.~\ref{feynman.fig} and will be discussed in the context of
transient bond formation. In this study where $T$ is high enough to prevent
strong bond formation, we use only the second order form.  Once the
pair-potential $V_{ii}(r)$ is determined, the ion-ion PDF $g_{ii}(r)$ can
be determined using a classical MD simulation; or via the   MHNC equation
coupled to the Ornstein-Zernike (OZ) equation. The later route, viz.,
NPA+MHNC is numerically very fast but requires a Bridge term $B(r)$ which
includes linked multi-ion correlations terms but not those arising
 from quantum effects like C-C bonding. 
\begin{eqnarray}
\label{MHNC.eq}
g(r)&=&\exp\{-\beta V_{ii}(r)+ h(r)-c(r)+ B(r)\} \\
\label{OZ.eqn}
h(r)&=& c(r)+\bar\rho\int d\vec{r}_1 h(\vec{r}-\vec{r}_1)c(\vec{r}_1)\\
h(r)&=& g(r)-1.
\end{eqnarray}
Thermodynamic consistency (e.g., the virial pressure being equal to the
thermodynamic pressure) is obtained by using the Lado-Foiles-Ashcroft
(LFA) criterion (based on the Bogoluibov bound for the free-energy) for
determining $B(r)$ using the hard-sphere model bridge
function~\cite{LFA83,chenlai91}. That is, the hard-sphere packing fraction
$\eta$ is selected according to an energy minimization that satisfies the
LFA criterion. The iterative solution of the MHNC equation, i.e.,
Eq.~(\ref{MHNC.eq}) and the OZ-equation, Eq.~(\ref{OZ.eqn}), yields a
$g_{ii}(r)$ for the ion subsystem. The LFA criterion and the associated
hard-sphere approximation can be avoided if desired, by using MD
 with the  pair-potential to generate the $g(r)$.

One may now ask if $\bar{\rho}g(r)$ should be used as the new ion-density
profile in Eq.~\ref{F-elecden.eqn} instead of the spherical cavity, as was
done in Ref.~\onlinecite{DWP1}. For all cases studied where the
bound-electron core is compactly contained inside the WS-sphere, such
further iterations are unnecessary. If the difference between the cavity
profile and the $g(r)$-profile is used in perturbation theory (or in a
Kohn-Sham calculation), the bound state distribution should not change
perceptibly, and hence $Z$ remains unchanged. The effect on $\Delta n_f(r)$
can be significant but the pseudopotential $U_{ei}(k)$ remains unchanged for
the conditions where a linear pseudopotential holds~\cite{cdw-Utah12}. This
is true since the effect of the  ion-density profile is subtracted out
before Eqs.~(\ref{pseudo.eq})-(\ref{vii.eq}) can be applied. The
thermodynamic consistency of the model is easily checked by
the satisfaction of the compressibility sum rule, viz.,
\begin{equation}
\label{comp.eq}
S_{ii}(k\to 0)=Z\bar{\rho}T\kappa_T,\; 
\kappa_T=-\frac{1}{v}\left[\frac{\partial V}{\partial p}\right]_T.
\end{equation}
Calculations confirm that carbon WDMs satisfy the requirements that enable
us to decouple the pair of DFT equations~\ref{F-elecden.eqn},
\ref{F-ionden.eqn} and use the NPA+MHNC mono-atomic equations for carbon
WDMs whose $\bar{\rho},T$ do not allow the formation of {\it stable}
covalent C-C bonds. If such molecular species are formed, then an explicit
multi-component NPA+MHNC formulation which includes molecular species,
i.e., a neutral-pseudomolecule (NPM) model has to be used. But a molecular
model is not needed and we closely recover available DFT+MD  and PIMC
results for carbon WDMs quite accurately. A comparison of $g(r)$ and the
density of states (DOS) from DFT+MD~\cite{galli89} and those of
 an NPA-like calculation for
carbon near the melting point were already presented in
1990~\cite{DWP-Carb90}.   

After such validation of the NPA, we use the method to uncover an unusual
phase transition in a somewhat lower-density carbon where there are also
only transient C-C bonds. This strongly-correlated carbon fluid at
$\simeq$7 eV  undergoes a liquid-to-gas type transition (`boiling'
transition) accompanied by a change in the state of ionization $Z$ of the
carbon atoms. The reduced ionization removes the electrons needed for
 C-C bonding and drives the transition. The  reduction in
the mean free-electron density causes a discontinuity in $Z$ with
$\Delta Z \simeq -1$.
\begin{figure}[t]
\includegraphics[width=.95\columnwidth]{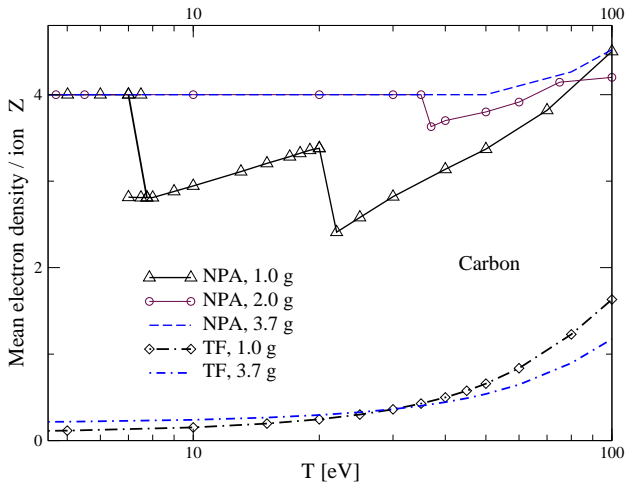}
\caption{The average ionization $Z$ of carbon as a function of $T$
for 1-3.7 g/cm$^3$. The ionization given by Thomas-Fermi (TF) theory,
More {\it et al.}~\cite{ZbarMore} is also given. Note the existence
of two possible values of $Z$ (hysteresis) from two possible Kohn-Sham
solutions near $T$=7 for the density 1.0 g/cm$^3$ at the phase-transition.
(see appendix). 
}
\label{zbar.fig}
\end{figure}
\subsection{The free-electron density
 per ion as an experimentally accessible quantity.}

The free-electron density per ion in a plasma, viz., $Z$, is even
claimed to ``not correspond to any well-defined observable in the
sense of quantum  mechanics'' by some workers, while some others regard
it as a `subjective, model dependent quantity'. Hence a
 discussion of $Z$  and a review of its discontinuities are useful. 

Unlike in simple quantum mechanics, quantum statistical theory requires
an associated heat bath and the introduction of quantities like  $T$,
$\mu$, $Z$, etc., which appear as Lagrange multipliers needed in the
theory~\cite{DWP1}, rather than simple operator mean values.
Alternatively, extended Hilbert spaces, thermofield dynamics etc., can be
used to give a completely  operator-like formulation or a hybrid
formulation for these quantities as well. These broader questions
are discussed briefly in Chapter 8 of Dharma-wardana~\cite{apvmm13}, and in
Heslot~\cite{mixedQ85}. Quantities like $Z$, i.e., the mean electron
density per ion, the chemical potential  $\mu$,  $T$, are
measurable physical quantities.

Ion-Sphere models confine the electron density $n(r)$ to the Wigner-Seitz
sphere, so that $n(r)=\bar{n}$ for $r>r_{ws}$.  The Kohn-Sham equation is
solved only within the WS-sphere. This results in a number of non-DFT
characteristics in IS models. The confinement is equivalent to the addition
of a spurious confinement potential $V_{conf}(r)$ via the imposed boundary
conditions. Hence $V_{conf}$ may be regarded as a non-local  one-body
potential. But DFT maps many-body interactions also to such a one-body
potential and hence the spurious $V_{conf}(r)$ can technically also
correspond to an interacting system. That is, the mapping to a
non-interacting {\it homogeneous} electron system with a chemical potential
$\mu^0$ is now replaced by a mapping to an inhomogeneous system  with a
$\mu_{conf}$ chosen so that the electrons inside the WS integrate to
$Z_n$.  This model is used in Purgatorio, MUZE etc,  but
Starette {\it et al.}~\cite{StaSauDalHam14} and
Piron {\it et al.}~\cite{PironBlenski11} have also used it in some of their
implementations, though not necessary. Figure 12 of Blenski {\it et
al.}~\cite{Blenski2013} shows  electron density oscillations outside
the ion-sphere, and yet we are told that the model fails for normal-density
Al as $T\to 0$, perhaps because the cavity is treated as an invariant
feature of their model which is not subtracted out. Given that the thermal
de Broglie wavelength of electrons is $\lambda_e\propto \sqrt{1/T}$,
confinement  becomes unphysical as $T\to 0$. In any case, any such
confinement is inconsistent with the free functional variation of
$\Omega(n,\rho)$ with $n(r)$ needed in Eq.~\ref{F-elecden.eqn}. The
non-variational nature of the IS-model was indeed noted by Blenski et al.
However, at sufficiently high $T$, the effect of  $V_{conf}(r)$ may be
neglected and the IS-approximation becomes valid. 

The correlation-sphere model, with $R_c=10 r_{ws}$ used in NPA does not
confine the electrons to the WS-sphere but solves the Kohn-Sham equation up
to $R_c$.  The central potential plus the neutralizing cavity is a weak scatterer,
where the free-electron
phase shifts (calculated at the $r\to R_c$ limit) satisfy the Friedel sum,
adding to zero.
 This procedure 
breaks down for bound-states which extend outside the WS-sphere.
 However, this is not the
case for the carbon plasmas studied here. 

There are several  specifications of $Z$ in IS
models~\cite{Murillo13,SternZbar07} but no self-consistency exists.
 Figure 4 of Stern {\it et
al.}~\cite{SternZbar07} display three  models of $Z$, labeled (a)
$Z_{continuum}$ which is the total number of unbound ($\epsilon_k>0$)
electrons in the WS-sphere (b) $Z_{backround}$, an estimate of the
continuum electron density using the ideal DOS, and (c) $Z_{ws}$, i.e., the
free-electron density on the WS-sphere surface, viz.,  $4\pi
(r_{ws})^2 n(r_{ws})/\bar{\rho}$. Since the IS model becomes
increasingly valid at high $T$, two of  the estimates, i.e.,  (b), (c) 
but not (a), approach each other as $T\to\infty$, but diverge as $T\to 0$.
  According
to Blenski {\it  et  al.}~\cite{Blenski2013} the $T\to 0$ estimate of $Z$
for normal-density aluminum obtained from their variational model 
approaches that of Thomas-Fermi theory (see Fig.~\ref{zbar.fig}). 
Refs.~\cite{eos95,PDWBenage02} discuss the aluminum EOS and $Z$
 via  NPA. For normal density Al, the NPA $Z$ equals 3 for $T<E_F$.
We do not support the view of Blenski {\it et al.}~\cite{Blenski2013}
that ``...all quantum models seem to give unrealistic description of atoms
 in plasma at low temperature and high plasma densities''.

One may ask what answers are available regarding $Z$ from more  detailed
simulations, e.g., DFT+MD methods using VASP, ABINIT and such codes. In
solid state physics, such codes provide electronic structure
data to calculate properties like
 the complex dielectric function~\cite{OptAl-Ehren63,
Seraphin71}, or specific Fermi surface  properties. These are used to
determine the free-electron density $\bar{n}$ per ion, i.e.,  $Z$.
 Even for a metal like
gold where the $d$-electrons extend outside the
WS-sphere, $\bar{n}$ per atom (i.e., $Z$)  contributing to the optical
conductivity is found to be unity for
$T<2$ eV, i.e., below the $5d\to 6s$ threshold, as is also found
from  optical experiments~\cite{Au-Theye72} on gold. Thus the
free-electron density per ion is a well-accepted quantity. This
view is current even in WDM studies (c.f., Review of XRTS phenomena  by Glenzer and
Redmer~\cite{GlenzerRedmer09}). 

Consider the experimental measurement of $Z$.  The
plasmon peak  positions (Stokes and anti-Stokes values)
 give a mean value of $\bar{n}$.  Their intensity ratio can
be used (assuming detailed balance) to obtain $T$ which is not an operator
but the Lagrange parameter
 associated with the conservation of $<H>$ where $H$ is the
Hamiltonian. Then, knowing the ion density $\bar{\rho}$ and $\bar{n}$, we
obtain $\bar{Z}$.  This $\bar{n}/\bar{\rho}$ must agree with the
$\bar{Z}$ measured independently from the conductivity which is obtained
from the absorbed, transmitted and reflected signals of a probe beam
yielding the dynamic conductivity $\sigma(\omega)$. The  $\omega\to 0$
limit of  $\sigma(\omega)$ yields the static conductivity and
hence $Z$. The XRTS ion-feature  $W(k)$ can be used to obtain another
estimate of $Z$. Hence we have at least three independent measures of $Z$
that the experimentalist obtains using (as far as possible) a single
self-consistent theory. Many other experimental measures of $Z$ exist
since $Z$  occurs in the ion-electron
pseudopotential, but we will not enumerate them here.  The NPA approach
offers  a consistent, parameter-free theory, as demonstrated in 
Ref.~\cite{xrt-LH16} and Ref.~\cite{cdw-plasmon16} for extracting $Z$ from
experimental data. DFT+MD  is also such a theory, although it does not yield $Z$
directly. Many WDMs and their
$T, Z$ and $\sigma$ have been  extracted using DFT+MD, from experiments using
femto-second laser probes. Even in space plasmas and discharge-tube
plasmas, $Z$, i.e., the free-electron density per ion, is measured
routinely using Langmuir probes, where the potential of the probe is
compared to a reference electrode. Hence, while  some scientists
 question the physical admissibility of `inderect'  properties like $Z,T$,
 we do not subscribe to that view. 
\subsection{The nature of the discontinuities in $Z$.}
\label{DisconZ}
Fig.~\ref{zbar.fig} reveals many discontinuities in $Z$ as a function of
$T$. We study in detail the discontinuity in $Z$ for $\rho\sim1$ g/cm$^3$
near $T=7$ eV and discuss the other discontinuities (e.g. at 22 eV, $\rho=$
1.0 g/cm$^3$) only briefly, leaving their detailed elucidation to future
work. The existence of such discontinuities in $Z$ has been well-known
and it has been a matter of concern for EOS studies.

The nature of the discontinuities in $\bar{Z}$ in NPA-like models needs to
be clarified. The one-atom NPA can be generalized to an $N$-atom
pseudo-molecule (NPM) calculation that becomes  analogous to the
$N$-atom quenched-solid model of WDM used in DFT+MD studies
via VASP, ABINIT etc. The $Z$
discontinuity problem in NPA becomes in NPM the well-known band-gap problem
in the DFT of solids, but now with {\it partial electron occupations} in
bands due to the finite-$T$ Fermi distribution. A limited solution to the
band gap problem at $T=0$  is to use the GW method in a somewhat
inconsistent but successful approximation without a vertex correction
$\Gamma(\omega)$. In the NPA, such a GW-like  implementation based on
solutions to the Dyson equation for H-plasmas  was given by Perrot and
Dharma-wardana in 1984~\cite{PDW-levelWidth}.  An alternative approach,
which has not been examined sufficiently by implementations is the use of
finite-$T$ ion-electron XC-potentials, self-interaction (SI) corrections etc., 
which may play a crucial role at these discontinuities.  In the case of
states which extend beyond the WS-sphere, some authors have tried  {\it ad
hoc} approaches using broadening or modifying the Kohn-Sham states.
But KS-states cannot carry any broadening as they apply to
 a {\it non-interacting} 
electron gas at the interacting density.

In Fig.~\ref{zbar.fig} we present the mean ionization $Z$ of isochoric
carbon at 3.7, 2.0 and 1.0  g/cm$^3$ as a function of $T$.   
Fig.~\ref{zbar.fig} indicates sharp {\it downward} change in $Z$, both for
C at  2.0 g/cm$^3$ and 1.0 g/cm$^3$. Unlike a rise in $Z$ caused by shell
ionization, downward drops are indicative of metal-semimetal  transitions.
If the bound-electron core of the carbon atom remains compactly within the
WS-sphere, and if the three estimates $Z_b, Z_F, Z_{cn}$ defined in
Eqs.~\ref{nb-Z.eq},\ref{Friedel.eq}, and Eq.~\ref{cn.eq} fall with 95\% 
of each other,  a converged NPA solution and its $Z$ are obtained.
 The deviations from self-consistency occur
  due to short-comings in $F_{xc}^{ee}$ which may lack SI
 corrections and multi-species effects.
The $\sim7$ eV fluid displays a transition between two ionization
species, $Z$=3 and $Z$=4, and even the AA prediction in the local density
approximation (LDA), without
SI corrections, allows $Z$ to dip to only 2.8 at
 $\sim7$ eV. SI corrections
would push $Z$ to 3. We show in Sec.~\ref{monomer-dimer.sec} that at 7 eV
the reduction in $Z$ is accompanied by a loss of ion correlations due to
loss of transient C-C bonding in the fluid.

If the $Z$ obtained from the NPA-LDA is a quantity like 2.5, then
model of a mixture of ions with
integral values of $Z=$ 2, 3, and 4  for C (as in Ref.~\cite{eos95} for
aluminum) is needed,  with an SI-corrected $F_{xc}^{ee}(n,T)$. Such corrections
 are also in DFT+MD studies of such transitions. Without a SI term, bound
states favour double occupancy, while the SI
(even a simple Hubbard $U$ model) favours single occupancy. Thus
 hydrogen-plasma models without SI contain only H$^-$ ions, with no H atoms.

{\it Discontinuities at higher $T$.}
No C-C bonding effects are likely even at low densities for $T>8$ eV as
$E_{cc}\simeq8$ eV even in dilute gases. But the discontinuity in $Z$ at 20 eV is very
similar to that at 7 eV.
The discontinuity in $Z$ at $\sim20-22$ eV for $\rho$=1.0 g/cm$^3$ is at
a $\Gamma_{ii}\simeq$2.26 while $T/E_F \simeq 2.5$, and  $Z\sim 2.5$. While
there is weak short-range order in the fluid,
the main effect is the re-population of the 2$p$ shell of carbon. The
 conductivity of the fluid calculated using 
LDA-XC drops, as in a metal-semimetal transition, but with
 no major change in the structure of the fluid with little ionic correlations.
  Given a mean $Z$ = 2.5, a description as mixture of electrons and ions, 
$Z_i= 2, 3,  4$, with compositions $x_i$, with suitable SI corrections applied to the
$F_{xc}^{ee}(n,T)$ is necessary to treat  this case (as in
the case of Al studied in Ref.~\cite{eos95}). Then a modest decrease
in the mean $Z$ and the conductivity may probably occur at 21 eV.
The slow rise of $Z$ from $\sim$ 2.5 at 22 eV towards 6 involves a
redistribution of the composition $x_i$ of ions over the long $T$-range.  

The discontinuity at 35 eV for carbon, $\rho$=2.0 g/cm$^3$ is of the same genre
as that for  carbon, $\rho$=1.0 g/cm$^3$ at T=21 eV. Here too the carbon fluid 
with $Z\simeq 3.6$ needs a treatment as a mixture of Z=3, and Z=4 species together
with XC-functional including SI, going beyond LDA.

{\it Discontinuities from quasibound states. --}
Discontinuities in $Z$  arise in LDA average-atom treatments of WDMs,
 especially in higher-$Z_n$
elements,  from bound states
emerging as quasi-bound states (QBS) extending
beyond $r_{ws}$.
Such QBS need not cause discontinuities in EOS properties due to compensating
changes in the DOS, e.g., as determined by the
phase-shifts of the continuum states~\cite{majumdar65}. 
 Several AA-models including that of Starrett and Saumon (SS)~\cite{StaSau13} use
a level-broadening parameter $\tau$ to smooth out QBS discontinuities in
$\bar{Z}$. As Kohn-Sham levels are
eigenstates of `non-interacting' electrons, a level width $\gamma$  is not
available within DFT where the Kohn-Sham states are for non-interacting electrons. 
If level widths are needed, a Dyson equation  has to be solved and the imaginary part of
the self-energy provides the level width, as  in
Ref.~\cite{PDW-levelWidth}.  Instead, a {\it transport} width  $1/\tau_{tr}$
obtained from the Ziman equation has been been used in
several  atom-in-plasma models  (e.g., see Appendix C of
Ref.~\cite{StaSau13} and references there in). The $1/\tau_{tr}$ is not a
 level width, and reflects
the momentum relaxation of continuum electrons in the $E_F\pm T$ window. The
QBS (e.g., $d$-bands calculated in an AA-model) arise from core-state features
lower in energy than the Fermi energy, and $1/\tau_{tr}$
cannot be simply transferred to other bands.
  In cases where
the bound states delocalize beyond the WS-sphere, a multi-center
`neutral-pseudo-molecular' formulation becomes necessary in DFT as constructing
suitable XC-functionals is difficult. Given that
valid discontinuities in $Z$ exist, forcing $Z$ and other properties to be
smooth across a discontinuity is to obliterate valid physics using an
artifice.

\subsection{NPA average ions and covalent bonding.}
Many low-$Z_n$ materials, e.g., H, O, N, P, Si and their mixtures show many
phases and covalent bonding  at low $T$. Such bonding persists into
their WDM states. The C-C bond in its various forms (single, double)  has a
bond energy $E_{\mbox{cc}}$  of the order of $\sim 4-8$ eV. However, in a
WDM medium, the presence of free electrons leads to screening effects absent
in non-conducting media, making the bonding weaker and transient at
higher temperatures $T\sim E_{\mbox{cc}}$. The valance $Z$ of carbon at low
$T$ even in nonconducting systems is 4, where three $p$ electrons and one
$s$-electron are available for hybridization. When bonding occurs, $n(r)$
 in the region {\it between} two carbon atoms (i.e., between
the two WS-spheres) increases, as easily seen even from the Heitler-London
description of bonding (sec. 7.3.2, Ref.~\cite{apvmm13}) and even
from the mid-bond density~\cite{PerRas83}. A spatial redistribution of
valance electrons occurs, but there is no change
 in $Z=Z_n-n_b$ as the core
electrons are not affected; the total number of free-electrons per atom is
conserved, and  $Z$ remains 4. No gap or pseudogap is found in the DOS near
 $E_F$ . An NPA-type 
 calculation for the DOS for carbon just above the melting point
 (transient C-C bonding), and a comparison with the
 corresponding DFT+MD
for the DOS and $g(r)$ of Galli {\it et al.} have been given in
 Ref.~\cite{DWP-Carb90}.  

 When transient bonding occurs, the relevant
configurations captured by the AA model of the NPA are an
 average over the transiently bonded ion
configurations. The pair-potential between these  average ions is given by
Eq.~\ref{vii.eq}.  The minima (or turning points) in $V_{ii}(r)$  can
produce `bonding' between NPA-carbons, while the principle minimum will
usually define a bond having the longest lifetime since the system is in a
thermal ensemble. The ion-ion PDFs in the non-bonding regime of densities
show peaks due to packing effects. As the bonding  region in the phase
diagram is approached, a prepeak appears. This is  an average over all
configurations around a single carbon ion. The bonding peak (`prepeak') in
$g(r)$ of liquid carbon has been analyzed in such terms by 
Correa {et al.}~\cite{CorreaBonev06}, Galli {\it et al.}\cite{galli89}
and by earlier workers as well.  
\begin{figure}[t]
\includegraphics[width=.95\columnwidth]{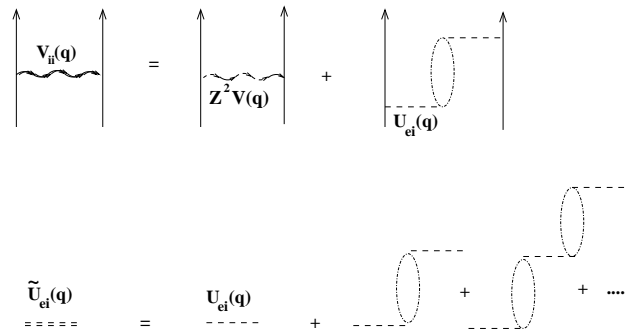}
\caption{Vertical arrow represent ion propagators, while the
electron-hole processes are shown as a loop. The direct ion-ion
Coulomb  interaction $Z^2V_q$ is converted
into the ion-ion pairpotential as in Eq.~\ref{vii.eq}. However,
the pseudopotential $U_{ei}(q)$ can be resumed to give $\tilde{U}_{ei}(q)
=U_{ei}/\{1-U_{ei}\chi(q)\}$ and indicates the possibility of
new collective modes associated with transient bond-formation in complex
liquids.}
\label{feynman.fig}
\end{figure} 

Thus an  ``average dimer'' in DFT is more complex than just two ions bonded
together. The ``effective bond'' that determines, say the first peak in
$g(r)$ is an average over  $sp3, sp2, sp$ configurations with one or many
neighbors. Even higher coordinations may occur at high compression. In the
NPA picture the shells of neighbours around an ion depend on the interplay
of many minima in the  Friedel oscillations in the ion-ion pair
potentials, mean-field effects and ion-correlation effects all
contributing to the  potential of mean force acting on a field ion.  At
very low $T$, the highly correlated C-C bonded forms of the liquid  become
increasingly favored, while at high $T$ the monomer-fluid  (i.e., single
ions) becomes dominant.  Thus, under favourable circumstances, a
`dimer-monomer' phase-transition can be anticipated as $T$ is increased.
The  valence electrons DOS modifies in forming a
transient band of bonding states  and anti-bonding states, but the mean
charge $Z$ (i.e., valance) does not change during bond-formation.
An increase of $T$ leads to
less screening and an increased nuclear attraction on electrons, binding
them, decreasing $Z$.
The available free electrons decrease and
 bonding decreases. Hence such `dimer-monomer'
transitions occur near a  {\it decrease} in $Z$.

The form of the pair-potentials implies that even the simplest NPA
approach implicitly allows a mixture of two fluids.  This view  is
theoretically equivalent to Stra\"{a}ssler and Kittel's~\cite{kittel65}
``pseudo-binary mixture'' model  basic to the theory of complex fluids. In
Ref.~\onlinecite{DWP-Carb90} Dharma-wardana and Perrot showed that the
ionic-structure of  liquid-metal phases of C, Si and Ge (which show unusual
shoulders at the mean peak)  can be obtained using an NPA-like approach. A
more recent discussion for WDM-Si is given in
Ref.~\onlinecite{cdw-Utah12}. The electrons form  transient bonds, and
become ``heavy'' due to  self-energy corrections accruing to the electron
propagators, specially near 2$k_F$.
 This Fermi-liquid model quantitatively explained the unusual
$S(k)$ of Si and Ge, viz., the shoulder on the high-$k$ side of the main
peak~\cite{DWP-Carb90,cdw-Utah12}  without invoking chemical
bonds  ``captured'' in computer simulations~\cite{galli89}. In the present
case the formation of time-dependent networks of C-C bonds  can be
understood as in Fig.~\ref{feynman.fig}. Neglecting vertex corrections and
ladder contributions, an  all-order interaction 
that extends over the liquid can be given as:
\begin{equation}
\label{utilde.eq}
\tilde{U}_{ei}(q)=U_{ei}(q)/\{1-U_{ei}(q)\chi(q)\} 
\end{equation}
The poles of  $\tilde{U}_{ei}(q)$ signal the
appearance of new elementary excitations in the liquid.
 The second-order form  becomes valid
only for  sufficiently high $T$ compared to the average energy of the C-C
bonds, as is the case for the 7 eV fluid studied here. The C-C bonded
ion-excitation is a break-off mode of the ion-optic folding of the
ion-acoustic mode leading to
complete localization in the ladder-diagram limit (not shown in the
figure). The pole-structure of Eq.~\ref{utilde.eq} shows that transient
C-C bond formation is a true collective phenomenon in complex conducting
liquids. Computer-simulation `snap shots' of quenched configurations
cannot capture its essentially dynamic Fermi-liquid  character.

\section {Comparisons with DFT+MD results, and PIMC simulations.}
\label{dft-pmic-npa.sec}
The applicability of the NPA for complex liquids was proposed back in
1990~\cite{DWP-Carb90}, but few DFT+MD calculations or  experiments for WDMs
existed to validate the NPA by detailed comparisons. Currently there are
several DFT+MD calculations, path integral Monte Carlo (PIMC) results, and
experiments;  we use them to compare the NPA calculations in this
section.
\subsection{Carbon at 3.7 g/cm$^3$ and at $\sim$100GPa.}
Experiments and DFT+MD  are available for carbon at a density of 3.7
g/cm$^3$,  $T\simeq$ 0.52 eV (6,000K)  and  0.86 eV (10,000K), with the
pressure within 100-200 GPa. The mean ionic charge $Z$ remains  4.0 even
beyond 50 eV while  $E_F$  is $29.9$ eV. Thus  WDM carbon probed by Kraus
{\it et al.}~\cite{kraus13} at 6,000K and 10,000K is such that $\theta=T/E_F
< 0.03$.  Hence the use of $T=0$ XC-functionals and  standard DFT+MD  is
justified. Kraus {\it et al.}  report the $S(k)$  obtained from their
simulations, while Whitley {\it et al.}~\cite{whitley15} have published 
$g(r)$ data for an overlapping regime. We use these results and the PIMC
predictions~\cite{driver12}   to benchmark the NPA method against DFT+MD as
well as PIMC methods.

\begin{figure}[t]
\includegraphics[width=.95\columnwidth]{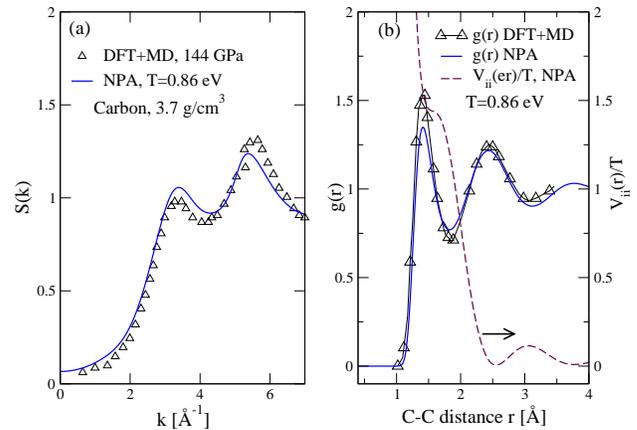}
\caption{(a) Comparison of the $S(k)$ of
liquid carbon from the NPA+HNC  with the DFT+MD (GGA) simulation
reported by Kraus {\it et al.}~\cite{kraus13} at 0.86 eV and 3.7g/cm$^3$. 
(b) The corresponding NPA C-C
potential (in units of $T$), 
the $g(r)$ from the NPA, and the DFT+MD $g(r)$  of
Whitley {\it et al.}~\cite{whitley15}.
}
\label{skp86ev-3p7.fig}
\end{figure}

 An all-electron Kohn-Sham calculation for a carbon nucleus immersed in an
electron fluid (with $\bar{n}$ = 0.1099 electrons per a.u. of volume when
the carbon density is 3.7 g/cm$^3$) inside a correlation sphere of radius of
10 $r_{ws}$ was used  to construct the neutral pseudo atom.  The
Wigner-Seitz radius is 2.0558 a.u.  The free-electron pile-up $\Delta
n_f(k)$ around the carbon nucleus defines $U_{ei}(k)$, and $V_{ii}(k)$,
i.e., Eq.~\ref{vii.eq}. The ion-ion structure factor $S(k)$ is calculated 
using the HNC equation and not the MHNC  for reasons discussed below. The
pair potential, $g(r)$ and  $S(k)$ for carbon at 3.7 g/cm$^3$, $T=0.86$ eV
are displayed in Fig.~\ref{skp86ev-3p7.fig}. Here the ion-ion coupling
constant $\Gamma_{ii}=Z^2/(r_{ws}T)\simeq$ 246 and hence we have a strongly
coupled fluid. Even with strong coupling, it is clear from 
Fig.~\ref{skp86ev-3p7.fig} panel (a) that the NPA+HNC calculation matches
the DFT+MD calculation quite closely.

\begin{figure}[t]
\includegraphics[width=.95\columnwidth]{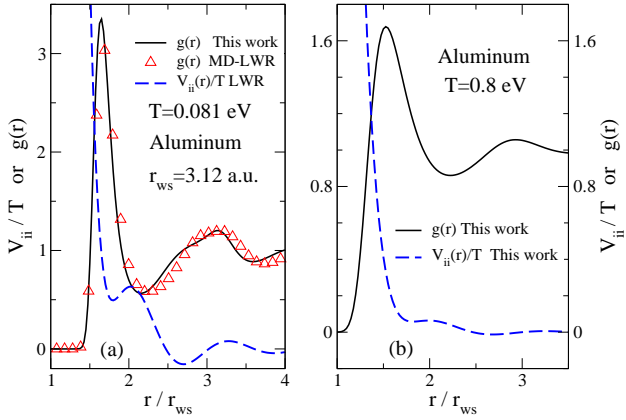}
\caption{(a) The pair-potential and $g(r)$ from the MD simulation of
normal-density Al at its melting point (0.081 eV) by 
Levesque-Weiss-Reatto (MD-LWR)~\cite{LWR85}
compared with our calculation. The main peak is near the {\it positive} energy
stationary point in the potential. The $g(r)$ peaks  roughly correspond to the 
 Friedel minima in the potential, and   phase out for larger $r$. 
(b) The corresponding 
$g(r)$ and pairpotential at a higher $T$, 0.8 eV, where the positive
stationary point in $V_{ii}(r)$ locates the main peak in $g(r)$,
as in Fig.~\ref{skp86ev-3p7.fig}(b).}
\label{alu.fig}
\end{figure}

The physical meaning of the peaks becomes clear by considering
 $S(k)$, and $g(r)$. In normal liquids (e.g., liquid
Aluminum at low $T$~\cite{cdw-aers83}), the first peak in $g(r)$
occurs near 1.6-1.7 times $r_{ws}$ (see Fig.~\ref{alu.fig});
 where as here the first peak is
$\simeq 1.3r_{ws}$, i.e., a short-range 'prepeak'  occurs earlier
 than the main peak of  simple dense fluids.

Such short-range correlations (large-$k$) involve bond formation and local
order. The hard-sphere bridge term was not designed for 
such bonding effects, and hence HNC is used rather than the MHNC. This
slightly underestimats the peak height, while the peak
positions are within $\sim$ 5\% of the DFT+MD peaks. The position of the
first peak ($\sim$ 1.4 \AA ) in $g(r)$ corresponds to a typical conjugated
C-C bond distance. From panel (b) of the figure we see that the C-C bond
formation corresponds to a stationary point in $V_{ii}$ with a {\it positive}
energy.  

Some readers may wonder if the turning point at a {\it positive}
energy in $V_{ii}(r)$ can be the cause of the first peak in $g(r)$.  As
cited in our 1990 study of carbon~\cite{DWP-Carb90}, this feature is familiar
from the $g(r)$ and the pair-potential of liquid aluminum (see
pair-potentials  given in Ref.~\cite{DWP-Carb90,HarbourCCP15}) and other
high-$Z$ liquids.  While the high-density electron fluid attempts to push the
ions inwards to become dense and gain the higher XC-energy of the
Gellman-Breuckner limit, the ions repel each other, and even the innermost
inflection point in $V_{ii}(r)$ becomes the best compromise. The MD
simulations for liquid aluminum near its melting point using the
 Dagens-Rasolt-Taylor potential
for aluminum  by Levesque {\it et al.} also demonstrate the same
features~\cite{CDW-LWR86,LWR85}, as shown in
 Fig.~\ref{alu.fig}(a). The same pseudopotential holds at a  higher
$T$, viz., 0.8 eV as $Z$ remains unchanged. The $T$=0.8 eV results are
shown in Fig.~\ref{alu.fig}(b).

 Bonding features (e.g., the pre-peak in the
$g(r)$ etc.) are unlikely to be captured by ion-sphere models
as they limit the electron pileup at the pseudo-ions to within the
ion-spheres, where as the C-C bonding requires an increased density in the
bonding regions {\it between the ions}. 
Starrett {\it et al.}~\cite{StaSauDalHam14} find no signs of C-C or C-H bonds
using their code, for CH-plastic in a regime where DFT+MD
studies show prepeaks due to C-C and C-H bonding.

The approximate correspondence of the successive peaks in $g(r)$ with
 the Friedel
oscillations in the pair-potential given in panel (b) should be noted. The
$g(r)$ is determined by the potential of mean force  $V_m(r)=V_{ii}(r)/T+$
Meanfield+Correlations, with $g(r)=\exp\{-V_m(r)\}$. 
The correspondence of the minima in $V_{ii}(r)$ and the maxima in $g(r)$
becomes less satisfactory for larger $r/r_{ws}$ when the mean field
and correlations kick in.
These systems cannot be modeled by
Thomas-Fermi or Yukawa-type potentials,  a dominant paradigm even
today. The DFT+MD simulations do not directly provide a pair-potential as
$N$-ions with $N\sim 100$ are used in the simulation. The DFT+MD $g(r)$
can in principle be inverted to yield a pair-potential. The inversion of a
$g(r)$ to a potential is not unique, and hence requires specifying a
 {\it first-principles form} for
$V_{ii}(r)$  and a $B(r)$, but it does not call for
dynamic structure data~\cite{chenlai91,March87,cdw-aers83}. One may also
proceed as in  Whitley {\it et al.}~\cite{whitley15}, who derived a C-C
potential at 0.86 eV  by force matching to DFT+MD. Their
extracted $V_{ii}$ (their Fig. 4(a)) is in good agreement with the
 NPA $V_{ii}(r)$, 
Fig.~\ref{skp86ev-3p7.fig}(b), but the  Friedel oscillations are hard to
capture using force-matching methods. 

\begin{figure}[t]
\includegraphics[width=.95\columnwidth]{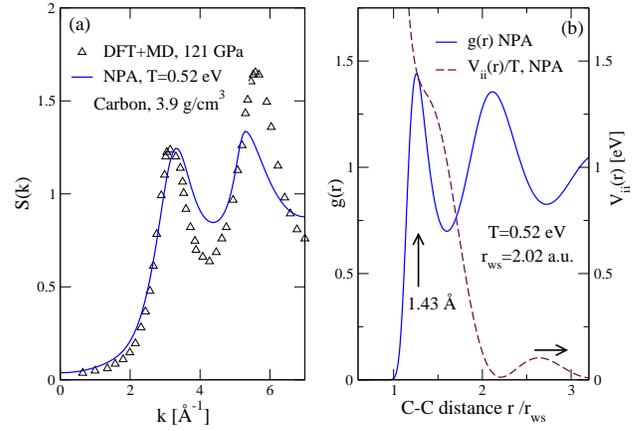}
\caption{(a) Comparison of the  $S(k)$ of liquid carbon from the NPA  with a
DFT+MD (LDA) simulation~\cite{kraus13} at 0.52 eV. (b) The corresponding NPA
C-C potential in eV, together with the $g(r)$ from the NPA+HNC.
}
\label{skp52ev-3p7.fig}
\end{figure}
 
 Applying  NPA+HNC to a very low $T$ case reveals its current limitations
as well as the limitations of the second-order pair-potential given by
Eq.~\ref{vii.eq} (see also Fig.~\ref{feynman.fig}). The DFT+MD simulations
of Kraus {\it et al}. at $\rho$=3.9 g/cm$^3$ and $T$ = 0.52 eV provide a
test case.  This temperature is an order of magnitude smaller than that of
the phase transition discussed in Sec.~\ref{monomer-dimer.sec} at $T=7$
eV, and the density is nearly four times higher, with $\Gamma_{ii}$=410. 
As there is C-C bonding at such low $T$ we use HNC (and not MHNC that
employs a hard-sphere bridge correction).   Figure~\ref{skp52ev-3p7.fig}
provides comparisons of NPA+HNC and DFT+MD at $T$=0.52 eV, i.e., 6000K.
The  $S(k)$ from NPA, Fig.~\ref{skp52ev-3p7.fig}(a) has  peak positions 
within $\sim$5\% of the DFT+MD $S(k)$, but the second peak height is
underestimated by as much as 19\%. This error in the peak height may due
to: (a) shortcomings in the 2nd-order pair-potential 
(c.f., Fig.~\ref{feynman.fig}),
 and  (b) the lack of
a suitable bridge function  for the C-C bonding peak in $g(r)$. These
issues will be treated elsewhere.

\subsection{Comparisons with PIMC.}
In astrophysics, much higher temperatures and densities occur. The C-C
association plays no role at the higher energy scales involved. Hence the
hard-sphere LFA bridge function should be applicable. Judging by the value
of $\Gamma_{ii}$,  NPA+MHNC should be accurate. Hence a comparison can
 mutually validate NPA as well as  PIMC calculations, and  also validate the
theory of $Z$ used in NPA. We calculate the $g(r)$ of liquid-carbon, density
12.64 g/cm$^3$,  at $T=8.62 $ eV (1x$10^5$K), 21.5 eV (2.5x10$^5$K), and
430.9 eV (5x10$^6$K) to compare with the PIMC and DFT+MD  calculations of
Driver and Militzer~\cite{driver12}.  The mean ionization $Z$ changes from
4.0 at low $T$, to 4.05 by $T=100$ eV, and reaches $Z$=5.758 by 430.9 eV,
the highest $T$ used by Driver {\it et al}. 

\begin{figure}[t]
\includegraphics[width=.95\columnwidth]{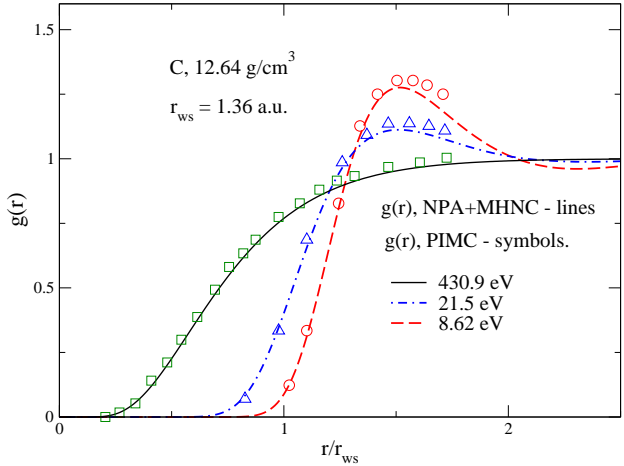}
\caption{ 
 The C-C pair-distribution functions $g(r)$ from the
NPA (lines), and from PIMC/DFT from Ref.~\cite{driver12}.}
\label{pimc.fig}
\end{figure}

 Fig.~\ref{pimc.fig} displays the $g(r)$ calculated using NPA+MHNC, and
PIMC or DFT+MD. The symbols show the PIMC or DFT+MD $g(r)$ from Ref.~\cite{driver12}
while the NPA+MHNC $g(r)$ are given as lines. The slight differences
 are due to limitations in the hard-sphere bridge
functions used in the MHNC, and due to limitations in the
calculations of Ref.~\cite{driver12} where only 24 carbon atoms have been
used in DFT+MD simulations. As also noted by Starrett and Saumon
(SS)~\cite{StaSau13} for $T$= 7.5$\times 10^5$ K (64.6 eV) given by
Driver {\it et al.}, the PIMC $g(r)$ does not tend to unity for large $r$,
but shows an error of $\sim$5\%. Our $g(r)$ for
7.5$\times 10^5$ K is not shown in Fig.~\ref{pimc.fig} to avoid overloading
the figure. But it  agrees closely with the $g(r)$ of SS
(Ref.~\cite{StaSau13}, Fig. 8) and the PIMC
data to the expected extent.  SS have not reported their $g(r)$ for lower $T$.

\begin{figure}[t]
\includegraphics[width=.95\columnwidth]{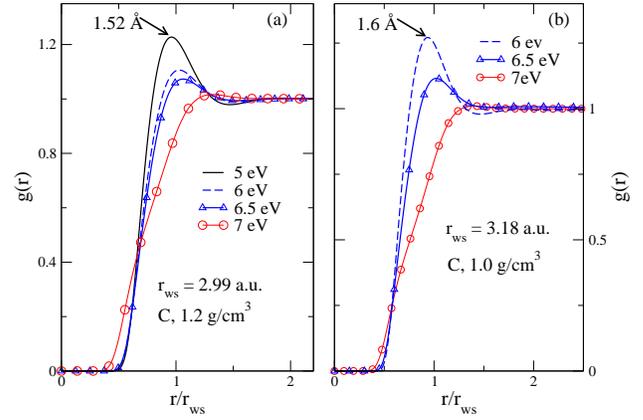}
\caption{(a) The evolution of the strongly correlated PDFs of
liquid-carbon, 1.2 g/cm$^3$ for $T<7$ eV across the 
(metal$\to$semi-metal)+(liquid $\to$ vapour) transition at $\simeq$ 7 eV
to $g(r)$ of a weakly correlated vapour. The 1.52 \AA$\,$ peak at 5 eV is
a signature of the C-C bond. (b) The evolution of  PDFs of liquid-carbon,
1.0 g/cm$^3$ across the transition. The C-C peak at $\sim$1.6 \AA$\,$,
T= 6 eV
is indicated.}
\label{gr1p2-1p0.fig}
\end{figure}

\section{Monomer-dimer phase-transitions in low-density liquid carbon.}
\label{monomer-dimer.sec}
An ionization and temperature (or pressure) driven plasma phase transition was
identified by Perrot and Dharma-wardana in WDM-aluminum via DFT~\cite{eos95}.
Such plasma phase-transitions had been proposed by Norman {\it et
al.}~\cite{Norman68} and by others on general grounds.  Simulations of such
transitions in WDM have been reported by, e.g., Redmer {\it et
al.}~\cite{LorHolRed10}. In a complex WDM liquid  like carbon, the novel
possibility of  dimer-monomer transitions which are driven by ``abrupt''
changes in the ionization $Z$ exists.  $Z$ must be 4 to have $sp3, sp2$ or
$sp$  bonded species in carbon. WDMs are high in energy-density 
and  $Z$-driven transitions are feasible.

Covalent bonding is enhanced at low densities and low temperatures. The
covalent C-C distance ranges from 1.3\AA$\;$ to $\sim$ 1.6\AA$\;$(3.02
a.u.) or even 1.7 \AA$\;$ depending on the coordination and screening by
free-electrons. When the density drops to less than  1.05 g/cm$^3$, the 
Wigner-Seitz radius $r_{ws}$ is 3.02 a.u., i.e., comparable to the C-C
distance.  We study liquid-carbon  at the density 1.0-1.4 g/cm$^3$ in the
context of changes in the mean ionization $Z$ with temperature, and the 
disruption of the `dimers' in favor of the monomers.

\begin{figure}[t]
\includegraphics[width=.95\columnwidth]{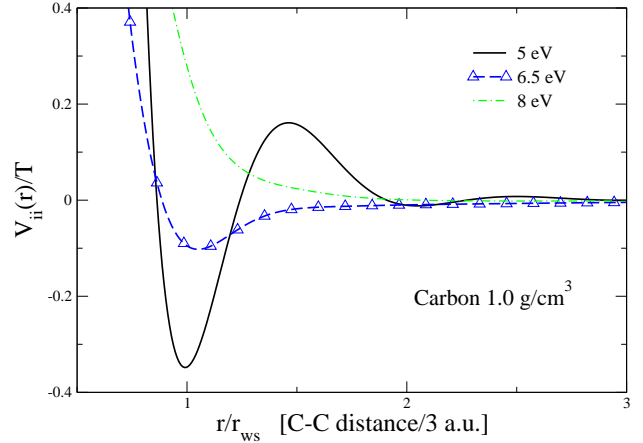}
\caption{Pairpotentials at $T$ =5, 6.5 and 8 eV for carbon WDM at 1.0
g/cm$^3$. At 5 eV $V_{ii}(r)$  is  negative ($-\sim$ 35\% of
the thermal energy) at its minimum. The minimum occurs at the C-C
bond distance of $\sim 3$ a.u. At this  $\rho,T$ carbon is  a
strongly correlated fluid. The potential at 6.5 eV yields the $g(r)$
shown in  Fig.~\ref{gr1p2-1p0.fig}(b) at 6.5 eV. At
8.0 eV $V_{ii}(r)$  is repulsive and the fluid becomes an
uncorrelated vapour.}
\label{pot5-8ev.fig}
\end{figure}

Fig.~\ref{zbar.fig} indicates sharp {\it downward} changes in $Z$, both for C
at  2.0 g/cm$^3$ and 1.0 g/cm$^3$. Unlike a rise in $Z$ caused by shell
ionization, downward drops are indicative of metal-semimetal  transitions.
Furthermore, the bound-electron core of the carbon atom remains compactly
within the WS-sphere, and hence there is no ambiguity in defining $Z=Z_n-n_b$
(this is further discussed in the appendix in the context of the transition).
Even in the transition region  the Friedel sum rule on $Z$, the $f$-sum rule,
and the conditions for the validity of the linear-response construction of
pseudopotentials (which depend on $Z$ being correct)~\cite{cdw-Utah12} are
satisfied to better than 93\%) by the NPA-LDA calculation. Hence the drop
in $\bar{n}$ and the discontinuity in $Z$ at 7 eV should be
observable in the EOS, $g(r)$, conductivity, compression etc., as further
 discussed below.

 In the case of carbon at $\sim$1.0 g/cm$^3$, an average `dimer'
fluid which has dynamical long-range order (Fig.~\ref{feynman.fig})
converts  to a monomer fluid. It is simultaneously 
a {\it metal$\to$semi-metal, liquid$\to$vapour}
transition, with a sharp drop in $Z$ at 7 eV for carbon at 1 g/cm$^3$. The 
properties show ``hysteresis'' typical of such transitions. 
 Since  $\Gamma_{ii}=Z^2/(r_{ws}T)$ for
 this case is $\simeq$ 19, NPA+HNC methods are eminently applicable here.

 In Fig.~\ref{gr1p2-1p0.fig} the $g(r)$ for two densities at the
transition are displayed for $5\le T\le7$ eV. At $T$=5 eV the mean charge
$Z$ is 4, while the  C-C distance as given by the 1st peak position in
$g(r)$, panel (a), is slightly larger than $r_{ws}$.   But it qualifies 
($\sim$ 1.52 \AA) for  a typical C-C bond with a bond length of 1.4-1.7
\AA. The peak weakens and the average bond length increases slightly as
the temperature is raised, as seen from the peak positions of the $g(r)$
at $T$=6 eV and 6.5 eV. Strikingly, at 7 eV all C-C correlations are
abruptly lost, and we have a mono-atomic carbon fluid with a  {\it
reduced}  ionization $Z \simeq$ 2.8. There is a gain in entropy in the
dimer $\to$ monomer transition, i.e.,  a complex liquid with transient C-C
bonding transiting to a random-monomer liquid, as in a liquid$\to$ vapor
transition. This transition clearly has analogies to the vaporization of
water where transient hydrogen bonds in liquid water break down
cooperatively to produced essentially water monomers in the vapour phase.
Hydrogen bonding in liquid-water involves  the lone-pair electrons in the
oxygen, and  its energy scale is much smaller  ($\sim 5$ kCal/mole) than
the transient C-C covalent bond relevant to the liquid-carbon phase
transition.

The evolution of the liquid from a strongly correlated C-C bonded
fluid to a disordered atomic vapour is manifested in the $V_{ii}(r)$
 shown in Fig.~\ref{pot5-8ev.fig}. Given the negative
pair-potential at 5 eV and below, persistent
C-C bonding occurs and the HNC equations do not
easily converge. However, NPA+MD can be used for such cases.

\begin{figure}[t]
\includegraphics[width=.75\columnwidth]{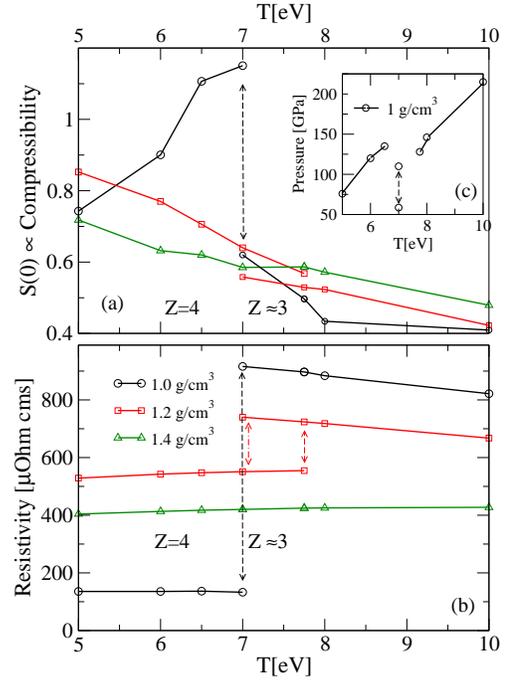}
\caption{Vertical arrows connect
two possible WDM-carbon fluids. 
(a) The variation of the compressibility of carbon 
for 1.0$\le\rho\le$1.4 g/cm$^3$ density and 5$\le T \le10$. (b) 
Variation of the  resistivity. (c) Variation of $P$ with $T$,
 $\rho=$ 1.0 g/cm$^3$. 
}
\label{res-comp-P.fig}
\end{figure}

Fig.~\ref{res-comp-P.fig} explores the critical region in $\rho$
(1.0-1.4 g/cm$^3$) and $T$ (5-10 eV). The transition occurs near
 $T=7$ eV with $\rho$ below $\sim$ 1.4 g/cm$^3$. At 1.4 g/cm$^3$
 $Z$ is 4, with no transition. For $\rho<1.4$, $Z$ dips to
 $\sim$ 3 near $T>$7 eV. The  compressibility $\kappa$ as reflected in 
$S(0)=\rho T \kappa$, Fig.~\ref{res-comp-P.fig}(a),  and the
resistivity $R$,  Fig.~\ref{res-comp-P.fig}(b), show discontinuities
 near 7 eV typical of a transition.

The pseudopotential $U_{ei}(k)$ and the structure factor $S(k)$ for carbon
obtained from the NPA+HNC calculation are used in the Ziman formula (used in
the form of Eq.~31,  Ref.~\cite{eos95}) to calculate the resistivity displayed
here.
The variation of the pressure with $T$, at $\rho=$ 1.0 g/cm$^3$
is shown in Fig.~\ref{res-comp-P.fig}(c) and confirms the transition. 
Furthermore,
for $\rho=1.2$ g/cm$^3$ the super-heating of the $Z=4$  fluid to
 7.75 eV, and the supercooling of the $Z\sim 3$ mono-ionic form 
from 7.75 to 7 eV give rise to two Kohn-Sham 
solutions.  Thus at 1.0 g/cm$^3$  a narrow metastable `two-solution'
 region around 7 eV exists. 

The NPA model accommodates a pseudo-binary mixture of fluids
 (e.g., monomer and average dimer) but cannot be 
pushed too far into lower $\rho, T$ since it treats the
fluid as a single `average' species and its {\it transient} pairs.
 For 6.5 eV $<T<$ 7.5 eV,
the value of $Z$ moves from 4 to approximately three. Since $Z$ is 
close to an integer value, the single-species AA
 model with LDA-XC is adequate. This is not the case near
 $T \ge 20$ eV where a
 multi-species C$^{+Z}$ fluid and XC-functionals beyond LDA are needed. 

 At lower densities and lower $T$, persistent molecular forms
 begin to dominate while monomers vanish.  Hence a more sophisticated
 `neutral-pseudo-molecule'
model  becomes necessary for applications to low-$T$ low-density complex
fluids with hardly any monomers. However,  carbon and other
low-$Z_n$ elements of the periodic table do not usually involve
interactions of bound electron states which extend outside their WS-spheres
in forming quasi-bound states
as in, say, gold. Hence 
many interesting low-$Z_n$ elements and their mixtures are likely to show
 phase transitions similar to
those of carbon.

{\it Conclusion. }
Calculations are presented to validate the NPA model for complex
liquids with transient covalent bonding, using the example of carbon. The method
recovers heavy DFT+MD and PIMC simulation results
with good accuracy from simple computations. The NPA 
 provides an unequivocal value for the
 mean-ionization $Z$, bound and free spectra, phase shifts,
pair-potentials, and structure factors needed in WDM calculations within a
single theory.  In regard to discontinuities in  $Z$, the need for
 XC-models which
 include GW and self-interaction corrections, and the need to use a mixture of
ions if $Z$ is close to a half-integer value, are emphasized.

Application to low-density liquid carbon has revealed a novel  
{\it a metal-semi-metal}  transition that occurs simultaneously with
 a {\it liquid-vapor} transition in the WDM regime. 
 Such transitions in carbon, (and most likely  in CH plastic
 used in ICF-ablators), and in other low-$Z_n$ elements
 may be of immediate importance in planetary physics,
 ICF ablator physics, and in carbon technologies.


\appendix*

\section{The electronic structure of Carbon at 6 eV to 10 eV, and the $Z$ discontinuity
at $\sim 7$ eV.}
The mean electron density/ion, $Z$, is defined in the NPA model as equal to
$Z_b=Z_n-n_b$, where $Z_n=6$, self-consistently agreeing with
 $Z_{cn}$, Eq.~\ref{cn.eq} and  with the Friedel sum $Z_f-Z=0$.
The bound electrons totaling $n_b$ should be  compactly contained in the
WS-sphere of carbon. That is, if the outermost bound-electron shell
has a mean radius $<r>$ distinctly less than $r_{ws}$, there is no ambiguity
in setting  $Z=Z_b$.  In fact $Z$ is the usual chemical valance of the
atom at normal density and pressure, but becomes non-integral for finite-$T$
where DFT-level occupations are Fermi factors, $f_{nl}$.  The
chemical potential used in $f_{nl}$ is $\mu^0$, the value for a {\it homogeneous
non-interacting} electron gas, consistent with DFT. The 
$T,\bar{n}$ dependence in $f_{nl}$, $\epsilon_{nl}$ and in the finite-$T$
phase shifts link  $Z_{cn}, Z_b, Z_F$ to  thermodynamics {\it via} the
 Kohn-Sham-Mermin formalism~\cite{eos95}. 
 The Kohn-Sham level structure and $Z$ at a carbon nucleus for
 liquid-carbon at
 1.0 g/cm$^3$
 at 6 eV-10 eV are given in Table~\ref{KSlevels.tab}. 
The rapid change in  $Z$ near 7 eV is
evident in Fig.~\ref{zbar.fig}.  The Kohn-Sham equation at
$T$=7 eV has two solutions, one with $Z$=4 and zero occupation
in the $n=2$ state, and another with $Z=2.78$ and having a
$n$=2 state with  an occupation of 
$\sim$0.71 and a mean radius $<r>=1.995$a.u. $Z$=2.78 is clearly
a $Z=3$ state corrupted by the lack of self-interaction corrections
in the $F_{xc}^{ee}$-LDA form.
 The bound state  is wholly inside the carbon Wigner-Seitz
sphere of radius 3.18 a.u. 
\begin{table}
\caption{Kohn-Sham energies $\epsilon_{nl}$ (a.u.) and occupations
 for carbon at 1.0 g/cm$^3$. Here 2$f_{nl}$ includes
spin. There are two solutions with two 
values of Z at $T$=7 eV. The bound-states with radii $<r>$ in a.u., 
are well inside the WS sphere,  $r_{ws}=3.18$ a.u. The Friedel
sum value of $Z$ from phase shifts, the charge-neutrality estimate 
$Z_{cn}=(4\pi/3)\bar{n}r_{ws}^3$ and $Z_b=6-n_b$
agree to better  than $\sim 93\%$ even at the phase transition.
When one value of $Z$ is given, agreement  
is $\sim$ 100\%. }
\begin{ruledtabular}
\begin{tabular}{lccccc}
 $T$ eV  &  $-\epsilon_{nl}$  &  2$f_{nl}$   & $<r>$ & $Z_{cn} $  \\
         &                    &              &       & $Z=6-n_b $ \\
\hline 
 6.0     &  $1s\;$,  19.181      &   2.00       & 0.269      & 4.0  \\
 7.0 (state 1)    &  $1s\;$,  18.647      &   2.00       & 0.273     & 4.0 \\
 7.0 (state 2)     &  $1s\;$,  18.932      &   2.00       & 0.273     &  2.78 \\
         &  $2s\;$,  0.1256      &   1.42       & 1.995     &  2.58\\
 8.0     &  $1s\;$,  19.793      &   2.00       & 0.267	     & 2.81 \\
         &  $2s\;$,  0.1709      &   1.33       & 1.854      & 2.67 \\

10.0     &  $1s\;$,  20.073      &   2.0        & 0.266      & 2.94 \\
         &  $2s\;$,  0.2196      &   1.15      & 1.757       & 2.85\\
\end{tabular}
\end{ruledtabular}
\label{KSlevels.tab}
\end{table}

Whitley {\it et al.}~\cite{whitley15}
used a constant mean ionization $Z$=3.2 (for the range of their study)
for carbon, together with  Yukawa screening
 to model the potentials fitted to their
 DFT+MD calculations at 0.86eV {\it via} a force-field analysis.
Their PDFs show features due to a C-C bond, as also 
recognized by them.
The existence of such bonds implies a value of $Z=4$
which is in fact compatible with their data.
The Thomas-Fermi ionization $Z_{tf}$ used in the conductivity
model of Lee and More~\cite{ZbarMore} (also in the  Quotidien
 equation of state (QEOS)~\cite{QEOS}) is displayed in  
Fig.~\ref{zbar.fig}(a) together  with the NPA $Z$ 
for liquid carbon in the range 1.0-3.7  g/cm$^3$. 
Thomas-Fermi (TF)  theory does not have shell structure, and
it is also not suitable for light elements at normal
and low densities.


\begin{thebibliography}{99}
\bibitem{galli89}
 G. Galli, 
R. M. Martin, R. Car, 
and M. Parrinello, Phys.
Phys. Rev. Let. {\bf 63}, 988 (1989).
 
\bibitem{glosli99}
 J. N. Glosli 
 and F. H. Ree, 
Phys. Rev. Let. {\bf 82}, 4659
(1999).

\bibitem{ghiring05}
L. M.  Ghiringhelli, Jan H. Los, A. Fasolino, 
Phys. Rev. B {\bf 72}, 214103, (2005).


\bibitem{DWP-Carb90}
M. W. C. Dharma-wardana and F. Perrot, Phys. Rev. Let., 
{\bf 65}, 76 (1990).

\bibitem{whitley15}
H. D. Whitley, 
 D. M. Sanchez , S. Hamel1, {\it et al.,}
Contrib. Plasma Phys. {\bf 55}, 390 (2015).

\bibitem{kraus13}
D. Kraus, 
J. Vorberger, D. O. Gericke, {\it et al.}
Phys. Rev. Let.  {\bf 111}, 255501 (2013).

\bibitem{hammelCH12}
Sebastian Hamel,
Lorin X. Benedict, Peter M. Celliers {\it et al.}
Phys.  Rev. B {\bf 86}, 094113 (2012).



\bibitem{hubNellis91}
W. B. Hubbard, 
W. J. Nellis, A. C. Mitchell, {\it et al.,}
Science {\bf 253}, 648 (1991).

\bibitem{sherman12}
B. L. Sherman, 
H. F. Wilson, D. Weeraratne, and B.Militzer, 
Phys. Rev. B {\bf 86}, 224113 (2012).

\bibitem{driver12}
K. P. Driver and B. Militzer,  Phys. Rev.
 Lett. \textbf{108}, 115502 (2012).


\bibitem{lindl04}
J. D. Lindl, 
P. Amendt, R. L. Berger, {\it et al.,}
Phys. Plasmas {\bf 11}, 339 (2004).

\bibitem{benedict2014}
L. X. Benedict, 
K. P. Driver, S. Hamel, {\it et al.}
Phys. Rev. B {\bf 89}  224109 (2014).

\bibitem{savvati08}
A. I. Savvatimskiy, Journal of Physics: Condensed Matter
  {\bf 20}, 114112 (2008).


\bibitem{losFaso03}
J. H. Los and A. Fasolino, Phys. Rev. B {\bf 68}, 024107 (2003).

\bibitem{losGhir05}
J. H. Los, 
 Luca M. Ghiringhelli, Evert Jan Meijer, and A. Fasolino,
Phys. Rev.  B {\bf 72}, 214102 (2005).

\bibitem{cdw-cpp15}
M. W. C. Dharma-wardana, 
Contrib. Plasma Phys. {\bf 55}, No.2-3, 79-81 (2015).

\bibitem{Pe-Be}
 F. Perrot,  Phys. Rev. E {\bf 47}, 570 (1993).

\bibitem{PDWBenage02}
F. Perrot, M. W. C. Dharma-wardana, and John Benage,
Phys. Rev. E {\bf 65}, 046414 (2002)


\bibitem{eos95}
F. Perrot and M.W.C. Dharma-wardana,
Phys. Rev. E. {\bf 52}, 5352 (1995).        
%
\bibitem{Murillo13}
Michael S. Murillo, 
 Jon Weisheit, Stephanie B. Hansen, and M. W. C. Dharma-wardana,
Phys. Rev. E {\bf 87}, 063113 (2013).

\bibitem{Muze05}
 G. Gregori, et al., Contrib. Plasma Phys. {\bf 45}, 284 (2005)


%
\bibitem{Purgatorio}
B. Wilson, 
V. Sonnad, P. Sterne, 
J. Quant. Spectrosc. Radiat. Transf. {\bf 99}  658-679 (2006).

\bibitem{StaSauDalHam14}
C. E. Starrett, D. Saumon, J. Daligault, and S. Hamel
Phys. Rev. E {\bf 90}, 033110 (2014)

\bibitem{Blenski2013}
T. Blenski, R. Piron, C. Caizergues and Bogdan Cichocki, High Energy Density Phys.
 {\bf 9}, 687-695 (2013)


\bibitem{PironBlenski11}
R. Piron and T. Blenski, Phys. Rev. E {\bf 83}, 026403 (201

\bibitem{SternZbar07} P.A. Sterne S.B. Hansen, B.G. Wilson,
 W.A. Isaacs, HEDP, {\bf 3}, 278 (2007)

%
\bibitem{DWP1}
M. W. C. Dharma-wardana and F. Perrot, Phys. Rev. A {\bf 26}, 2096  (1982)

\bibitem{ilciacco93}
E. K. U. Gross, and R. M. Dreizler,
{\it Density Functional Theory},
 NATO ASI series, {\bf 337}, 625
 Plenum Press, New York (1993).

%



\bibitem{Chihara87}
J. Chihara, J. Phys. F: Met. Phys. {\bf 17} 295 (1987)

\bibitem{Furutani90}
F. Perrot, Y. Furutani and M.W.C. Dharma-wardana,
Phys. Rev. A {\bf 41}, 1096-1104 (1990)

%

%
\bibitem{Dagens2}
L. Dagens, J. Phys. (Paris) \textbf{36}, 521 (1975).


%
\bibitem{HarbourCCP15}
L. Harbour, M. W. C. Dharma-wardana, D. D. Klug, L. J. Lewis,
Contr. Plsma. Phys.  {\bf 55}, 144-151 (2015)

\bibitem{GlenzerRedmer09}
S. H.  Glenzer and Ronald  Redmer, Rev. Mod. Phys. {\bf 81} 1625 (2009) 

%
\bibitem{PDW-levelWidth}
F. Perrot and M. W. C. Dharma-wardana, Phys. Rev. A {\bf 29}, 1378 (1984)
%


\bibitem{cdw-Utah12}
M. W. C. Dharma-wardana, 
Phys. Rev. E {\bf 86}, 036407 (2012).
%
\bibitem{xrt-LH16}
L. Harbour, M. W. C. Dharma-wardana, D. D. Klug and L. J. Lewis, 
Physical Review E {\bf 94}, 053211, (2016) 

\bibitem{PDWXC}
F. Perrot and M. W. C. Dharma-wardana, Phys. Rev. B {\bf 62}, 16536 (2000);
{\it Erratum: } {\bf 67}, 79901 (2003); arXive-1602.04734 (2016).

\bibitem{LFA83}
F. Lado, S. M. Foiles, and N. W. Ashcroft,
Phys. Rev. A {\bf 28}, 2374  (1983).

\bibitem{chenlai91}   
H. C. Chen and S. K. Lai,
Phys. Rev. A {\bf 45}, 3831 (1992).

\bibitem{apvmm13}
M. W. C. Dharma-wardana, {\it A physicist's view of
matter and mind}, Ch 8-9, World Scientific, New Jersey (2013)

\bibitem{mixedQ85}
A. Heslot, Phys. Rev. D  {\it 31}, 1341 (1985)

\bibitem{OptAl-Ehren63}
H. Ehrenreich, H. R. Philipp, AND B. Segal, Phys. Rev. {\bf 132},
 1918 (1963)

\bibitem{Seraphin71}
N. E. Christensen and B. O. Seraphin, Phys. Rev.
B {\bf 4}, 3321 (1971).

\bibitem{Au-Theye72}
M. L. Th\`{e}ye, Phys. Rev. B {\bf 2}, 3060 (1970)


\bibitem{cdw-plasmon16}
M. W. C. Dharma-wardana,
Phys. Rev. E {\bf 93}, 063205 (2016);
arXiv:1602.04734 (2016).

\bibitem{March87}
N. H. March, Can. J. of Phys., {\bf 65}, 219-240, (1987).


\bibitem{cdw-aers83}
M.W.C. Dharma-wardana 
and G.C. Aers,  
Phys. Rev. B. {\bf 28}, 1701 (1983).


\bibitem{majumdar65}
 W. Kohn and C. majumdar, Phys. Rev. {\bf 138}, A1617 (1965)

\bibitem{StaSau13}
C. E. Starrett and D. Saumon, Phys. Rev. E {\bf 87}, 013104 (2013)

\bibitem{PerRas83}
F. Perrot and M. Rasolt, Phys. Rev. B {\bf 27}, 3273 (1983)


\bibitem{CorreaBonev06}
Alfredo A. Correa, 
Stanimir A. Bonev, and Giulia Galli,
PNAS, {\bf 103}, 1204 (2006)


\bibitem{kittel65}
S. Strassler and C. Kittel, Phys. Rev. {\bf 139}, A758 (1965).


\bibitem{CDW-LWR86}
Dharma-wardana, M. W. C. and Aers, G. C.,
 Phys. Rev. Lett.{\bf 56}, 1211 (1986)

\bibitem{LWR85}
D. Levesque, J. J. Weis, and L. Reatto
Phys. Rev. Lett. {\bf 54}, 451  (1985).


\bibitem{Norman68}
G. E. Norman 
 and A. N. Starostin,
 High Temp. {\bf 6}, 394 (1968).

\bibitem{LorHolRed10}
W. Lorenzen, 
B. Holst, and R. Redmer, 
Phys. Rev. B {\bf 82}, 195107 (2010).



\bibitem{ZbarMore}
Y. T. Lee and R. M. More, Phys. Fluids {\bf 27}, 1273 (1984).

\bibitem{QEOS}
R.M. More,
 K.H. Warren, D.A. Young, and G.B. Zimmerman, 
Phys. Fluids {\bf 31}, 3059 (1988).















\end{thebibliography}
\end{document}